\documentclass[a4paper,epsfig]{jpconf}
\usepackage{graphicx}
\def \lleq {\lower0.9ex\hbox{ $\buildrel < \over \sim$} ~}
\def \ggeq {\lower0.9ex\hbox{ $\buildrel > \over \sim$} ~}

\def \om    {\Omega}

\def \obh {\Omega_b h^2}
\def \omms   {\Omega_m}

\def \omm  {\Omega_{0 {\rm m}}}
\def \oml   {\Omega_{\Lambda}}

\def \beq  {\begin{equation}}
\def \eeq  {\end{equation}}
\def \ber  {\begin{eqnarray}}
\def \eer  {\end{eqnarray}}

\newcommand{\newc}{\newcommand}
\newc{\be}{\begin{equation}}
\newc{\ee}{\end{equation}}
\newc{\ba}{\begin{eqnarray}}
\newc{\ea}{\end{eqnarray}}
\newc{\bea}{\begin{eqnarray*}}
\newc{\eea}{\end{eqnarray*}}
\newc{\D}{\partial}
\newc{\ie}{{\it i.e.} }
\newc{\eg}{{\it e.g.} }
\newc{\etc}{{\it etc.} }
\newc{\lcdm }{$\Lambda$CDM }
\newcommand{\nn}{\nonumber}
\newc{\ra}{\rightarrow}
\newc{\lra}{\leftrightarrow}
\newc{\gsim}{\buildrel{>}\over{\sim}}
\newcommand {\ga} {\ {\raise-.5ex\hbox{$\buildrel>\over\sim$}}\ }
\newcommand {\la} {\ {\raise-.5ex\hbox{$\buildrel<\over\sim$}}\ }
\newcommand{\eqn}[1] {Eq.~(\ref{#1})}

\renewcommand{\(}{\left(}

\renewcommand{\[}{\left[}
\renewcommand{\]}{\right]}
\begin{document}
\title{Consistency of $\Lambda$CDM with Geometric and Dynamical Probes}

\author{L. Perivolaropoulos}

\address{Department of Physics, University of Ioannina, Greece}

\ead{leandros@uoi.gr}

\begin{abstract}
The $\Lambda$CDM cosmological model assumes the existence of a small cosmological constant in order to explain the observed accelerating cosmic expansion. Despite the dramatic improvement of the quality of cosmological data during the last decade it remains the simplest model that fits remarkably well (almost) all cosmological observations. In this talk I review the increasingly successful fits provided by $\Lambda$CDM on recent geometric probe data of the cosmic expansion. I also briefly discuss some emerging shortcomings of the model in attempting to fit specific classes of data (eg cosmic velocity dipole flows and cluster halo profiles). Finally, I summarize recent results on the theoretically predicted matter overdensity ($\delta_m=\frac{\delta \rho_m}{\rho_m}$) evolution  (a dynamical probe of the cosmic expansion), emphasizing its scale and gauge dependence on large cosmological scales in the context of general relativity. A new scale dependent parametrization which describes accurately the growth rate of perturbations even on scales larger than $100h^{-1}Mpc$ is shown to be a straightforward generalization of the well known scale independent parametrization $f(a)=\omms(a)^\gamma$ valid on smaller cosmological scales.
\end{abstract}

\section{Introduction}
Converging geometrical \cite{scp,hzsst,Hicken:2009dk,cmb,bao,lpgeom}  and dynamical \cite{Tegmark:2003ud,dynprob,Nesseris:2007pa,Polarski:2007rr,lindercahn} cosmological observations indicate that the universe has entered a phase of accelerating expansion. This expansion could be driven either by the negative pressure of a homogeneous dark energy component \cite{dde} or by a modified gravitational force which is repulsive on large cosmological scales. Alternatively it could simply be due to an unusually large underdensity (void) on scales of about 1Gpc which induces an apparent isotropic accelerating expansion to observers located approximately at the center of the void \cite{void}. All of the above three classes of models attempting to explain the observed accelerating expansion suffer from fine tuning problems which are reflected in the unexpected values of the parameters required to fit the cosmological expansion data. For example dark energy originating from an evolving scalar field requires a scalar field mass of order $10^{-42}GeV$ which is several orders of magnitude smaller from expectations based on particle physics theories. Similarly, models assuming the existence of large voids require fine tuning of the location of the earth based observer within about $20Mpc$ from the $1Gpc$ large void while the formation of such a large void is also very unlikely in the context of standard cosmology \cite{void}.

The simplest and most successful model consistent with (almost) all cosmological  data is the $\Lambda$CDM model\cite{lcdmrev}. This model is based on a modified version of Einstein's equations of the form \beq R_{\mu \nu}-\frac{1}{2} R g_{\mu\nu} + \Lambda g_{\mu \nu} = 8 \pi G T_{\mu\nu} \label{eeql}\eeq where $\Lambda$ is a new free parameter, the cosmological constant. When the new term $\Lambda g_{\mu \nu}$ is placed on the left hand side (lhs) of equation (\ref{eeql}) it is interpreted as a modification of the gravitational law corresponding to an effective Newton potential \cite{lcdmrev} \beq V(r)=-\frac{GM}{r}-\Lambda r^2 \label{ptll} \eeq where $-\Lambda r^2$ corresponds to a new repulsive gravitational term. When the new term $\Lambda g_{\mu \nu}$ is placed on the right hand side (rhs) of equation (\ref{eeql}) it is interpreted as a new contribution to the energy momentum tensor $T_{\mu \nu}$ corresponding to dark energy with constant density $\rho_{\Lambda}=\frac{\Lambda}{8\pi G}$ and constant negative pressure $p_\Lambda = -\rho_\Lambda$.

Consistency with cosmological data requires a fine tuned value for $\rho_{\Lambda}$ \beq  \rho_{\Lambda}^{(obs)}\simeq (10^{-12}GeV)^4\simeq 2\times 10^{-10} erg/cm^3 \label{rlobs}\eeq
A physically motivated origin of the cosmological constant could be the energy of the quantum field vacuum which is predicted to have a constant diverging energy density and negative pressure. Under the assumption of a proper cutoff the quantum vacuum can be made finite and play the role of a cosmological constant. A natural scale for this cutoff is the Planck scale leading to a vacuum density \beq  \rho_{\Lambda}^{(Pl)}\simeq (10^{18}GeV)^4\simeq 2\times 10^{110} erg/cm^3 \label{rlpla}\eeq which is 120 orders of magnitude larger than the observed value.

Despite of the unnaturally small value of the observed $\rho_{\Lambda}^{(obs)}$ when compared to the anticipated value $\rho_{\Lambda}^{(Pl)}$ in the context of vacuum energy, the cosmological constant of equation (\ref{eeql}) has some unique physically motivated features. In particular the lhs of equation (\ref{eeql}) is the most general second rank tensor which is \begin{itemize} \item local \item coordinate covariant \item divergenceless (needed for energy momentum conservation) \item symmetric \end{itemize} An additional important attractive feature of $\Lambda$CDM is simplicity. It is the only model based on General Relativity (GR) which (assuming flatness) involves a single free parameter ($\oml\equiv \frac{\rho_{\Lambda}}{\rho_c}= 1-\omm$ where $\rho_c$ is the present day critical density required for a flat universe) and its predicted expansion rate $H(z)=\frac{\dot a}{a}(z)$ as a function of redshift $z$ \beq H(z)^2=H_0^2 \[\omm (1+z)^3 +(1-\omm)\] \label{hzl} \eeq is currently consistent with all cosmological expansion probes (geometric and dynamical). It is therefore clear that the identification of potential conflicts of $\Lambda$CDM with cosmological data is a prerequisite before seriously considering alternative more complicated models unless such models are free from any type of fine tuning. Unfortunately no such model is currently known.

Cosmological observations testing the $\Lambda$CDM model include two classes of probes \begin{itemize} \item Geometric probes which measure directly the cosmic metric (eg Type Ia supernovae as standard candles or Cosmic Microwave Background (CMB) spectrum peaks and Baryon Acoustic Oscillations (BAO) which use the last scattering horizon scale as a standard ruler) \item Dynamical probes which measure simultaneously the cosmic metric and the gravitational law on cosmological scales. The main dynamical probe is the evolution of linear dark matter cosmological perturbations $\delta_m(k,z)=\frac{\delta \rho_m}{\rho_m}(k,z)$ as a function of scale and redshift. This evolution can be probed either directly through weak lensing observations\cite{weak-lens} or indirectly though the large scale power spectrum of luminus matter at various redshifts \cite{Nesseris:2004wj}. \end{itemize}
Geometric probes provide currently the most accurate information about the cosmic expansion rate $H(z)$ as a function of redshift. In the first part of this brief review (section 2) I discuss the following three basic questions \begin{enumerate} \item What is the figure of merit \cite{tfor} (constraining power) of various recent Type Ia supernovae (SnIa) datasets and how does it compare with the corresponding figure of standard ruler (CMB+BAO) geometric probe data? \item What is the consistency level of various geometric probe datasets with $\Lambda$CDM? \item What is the level of consistency between standard candle SnIa datasets and standard ruler CMB+BAO geometric probes? \end{enumerate}
A few cosmological observations which appear to be in some tension \cite{Perivolaropoulos:2008ud} with the predictions of $\Lambda$CDM will also be discussed in section 2. The most interesting of these observations appears to be the observed dipole velocity flows on scales larger than $50h^{-1} Mpc$ which appear to be a factor of about four larger than the $\Lambda$CDM model predictions assuming normalization of the matter fluctuation power spectrum using the WMAP5 CMB data\cite{Watkins:2008hf,Kashlinsky:2008ut,Feldman:2009es}. Other puzzling observations include the high redshift brightness of SnIa \cite{Perivolaropoulos:2008yc}, the cluster halo profiles \cite{Broadhurst:2008re,Broadhurst:2004bi} and the emptiness of voids \cite{Tikhonov:2008ss,Peebles:2003pk,Klypin:1999uc}.

The theoretically predicted linear evolution of cosmological perturbations $\delta_m(k,z)$ as a function of redshift and scale, depends sensitively on both the expansion rate and the gravitational law on cosmological scales. This dual sensitivity makes the linear evolution of density perturbations a particularly useful dynamical cosmological probe. Even though current observational constraints on $\delta_m(k,z)$ are not as powerful as geometric constraints on $H(z)$ this is expected to change dramatically in the next decade \cite{futde}.

The time evolution of matter density perturbations $\delta_m(k,t)$ is described by the well known equation (see eg \cite{dynprob}) \beq {\ddot \delta_m} + 2 H {\dot \delta_m} - 4\pi G \rho_m f(k,t) \delta_m =0 \label{greqtim} \eeq where in the context of general relativity $f(k,t)=1$ and $\delta_m(k,t)=\delta_m(t)$ becomes independent of the scale $k$. In fact, this scale independence of $\delta_m$ is occasionally considered to be a signature of validity of general relativity\cite{Acquaviva:2008qp}. However, the derivation of equation (\ref{greqtim}) is based on two important assumptions whose validity is questionable on scales $(\gsim 200 h^{-1} Mpc)$. These assumptions are the following: \begin{enumerate} \item The scale of the perturbation $\delta_m$ is significantly smaller than the Hubble scale {\it at all} times during the perturbation evolution. \item The choice of gauge plays a minor role in the form of the evolution equation for $\delta_m$. \end{enumerate} An important question discussed in the second part of this review (section 3) is the following: {\it What is the level of validity of these assumptions on large cosmological scales?} It can be shown that the validity of both assumptions breaks down rapidly as the scale increases beyond $200 h^{-1} Mpc$.

\section{$\Lambda$CDM confronts Recent Geometric Cosmological Data}

A useful way to test the \lcdm model is to consider a generalized parametrization of the dark energy equation of state which includes \lcdm as a special parameter case and find the likelihood of the parameters corresponding to \lcdm in this context. A commonly used such parametrization is the Chevallier-Polarski, Linder (CPL)\cite{Chevallier:2000qy,Linder:2002et} ansatz \be w =w_0+w_1
(1-a)=w_0+w_1\frac{z}{1+z} \label{cpl} \ee which reduces to \lcdm for $(w_0,w_1)=(-1,0)$.
In this section, I show the ranking of the six latest Type Ia supernova (SnIa) datasets (see Table \ref{tabsnsets})
(Constitution (C), Union (U), ESSENCE (Davis)  (E), Gold06 (G),
SNLS 1yr (S) and SDSS-II (D)) in the context of the
CPL parametrization (\ref{cpl}) according to their Figure of Merit (FoM)\cite{tfor}, their
consistency with the cosmological constant ($\Lambda$CDM) and their
consistency with standard rulers (Cosmic Microwave Background
(CMB) and Baryon Acoustic Oscillations (BAO)). The datasets considered are shown in Table (\ref{tabsnsets}) along with some useful features
such as the redshift range or the subsets of each set\cite{Sanchez:2009ka}.

Assuming a CPL parametrization for
$w(z)$ (equation (\ref{cpl})) it is possible to apply the maximum likelihood
method separately for standard rulers (CMB+BAO) and standard
candles (SnIa) assuming flatness. The corresponding late time form
of $H(z)$ for the CPL parametrization is \ba H^2 (z)&=&H_0^2 [
\Omega_{\rm 0m} (1+z)^3 + \nn
\\ &+& (1-\Omega_{\rm 0m})(1+z)^{3(1+w_0+w_1)}e^{\frac{-3w_1
z}{(1+z)}}]\,. \label{hcpl} \ea

\begin{table}[h]
\caption{The datasets used in the present analysis. See respective
references for details on the sources of the SnIa data points.
\label{tabsnsets}}
\begin{center}
\begin{tabular}{ccccc}
\br
\textbf{Dataset} \hspace{7pt}& \textbf{Date Released}\hspace{7pt}& \textbf{Redshift Range} \hspace{4pt} & \textbf{\# of SnIa}\hspace{5pt} & \textbf{Filtered subsets included}\hspace{7pt} \\
\mr
SNLS1 \cite{Astier:2005qq}      & 2005 \hspace{7pt} & $0.015\leq z \leq 1.01$ \hspace{4pt} & 115  \hspace{7pt} & SNLS \cite{Astier:2005qq}, LR \cite{lr}                                                   \\
\vspace{6pt} Gold06 \cite{Riess:2006fw}     & 2006 \hspace{7pt} & $0.024\leq z \leq 1.76$ \hspace{4pt} & 182  \hspace{5pt} & \begin{tabular}{c} SNLS1 \cite{Astier:2005qq}, \\ HST \cite{Riess:2006fw}, SCP \cite{scp}, \\ HZSST \cite{hzsst} \end{tabular}  \\
\vspace{6pt} ESSENCE  \cite{WoodVasey:2007jb},\cite{Davis:2007na}   & 2007 \hspace{7pt} & $0.016\leq z \leq 1.76$ \hspace{4pt} & 192  \hspace{5pt} & \begin{tabular}{c} SNLS1 \cite{Astier:2005qq}, HST \cite{Riess:2006fw},\\ ESSENCE\cite{WoodVasey:2007jb},\cite{Davis:2007na}\end{tabular}       \\
\vspace{6pt} Union  \cite{Kowalski:2008ez}      & 2008 \hspace{7pt} & $0.015\leq z \leq 1.55$ \hspace{4pt} & 307  \hspace{5pt} & \begin{tabular}{c} Gold06 \cite{Riess:2006fw}, ESSENCE\cite{WoodVasey:2007jb}, \\ \cite{Davis:2007na} \end{tabular}  \\
\vspace{6pt} Constitution \cite{Hicken:2009dk} & 2009 \hspace{7pt} & $0.015\leq z \leq 1.55$ \hspace{4pt} & 397  \hspace{5pt} & Union \cite{Kowalski:2008ez}, CfA3\cite{Hicken:2009dk} \\
\vspace{6pt} SDSS  \cite{Kessler:2009ys}   & 2009 \hspace{7pt} & $0.022\leq z \leq 1.55$ \hspace{4pt} & 288  \hspace{5pt} & \begin{tabular}{c} Nearby \cite{fitters}, SDSS-II \cite{Kessler:2009ys},\\  ESSENCE \cite{WoodVasey:2007jb}, SNLS \cite{Astier:2005qq}, \\ HST \cite{Riess:2006fw}\end{tabular}   \\
\br
\end{tabular}
\end{center}
\end{table}

The SnIa observations use light curve fitters\cite{fitters} to
provide the apparent magnitude $m(z)$ of the supernovae at peak
brightness. The resulting apparent magnitude $m(z)$ is related to
the dimensionless luminosity distance $D_L(z)$ through \cite{Lazkoz:2007cc,Sanchez:2009ka} \be
m_{th}(z)={\bar M} (M,H_0) + 5 log_{10} (D_L (z))\,, \label{mdl}
\ee
where
\be D_L (z)= (1+z) \int_0^z
dz'\frac{H_0}{H(z';\om,w_0,w_1)} \label{dlth1} \ee is the Hubble
free luminosity distance ($H_0 d_L$),  and ${\bar M}$ is the
magnitude zero point offset and depends on the absolute magnitude
$M$ and on the present Hubble parameter $H_0$ as \ba
{\bar M} &=& M + 5 log_{10}(\frac{H_0^{-1}}{Mpc}) + 25= \nn \\
&=& M-5log_{10}h+42.38 \label{barm}. \ea The parameter $M$ is the
absolute magnitude which is assumed to be constant after proper corrections (using light curve fitters) have been implemented in $m(z)$.

The theoretical model parameters are determined by using the maximum likelihood method ie by minimizing the
quantity \be \chi^2_{SnIa} (\om,w_0,w_1)= \sum_{i=1}^N
\frac{(\mu_{obs}(z_i) - \mu_{th}(z_i))^2}{\sigma_{\mu \; i}^2 }
\label{chi2} \ee where $N$ is the number of SnIa of the dataset
and $\sigma_{\mu \; i}^2$ are the errors due to flux
uncertainties, intrinsic dispersion of SnIa absolute magnitude and
peculiar velocity dispersion. These errors are assumed to be
Gaussian and uncorrelated. The theoretical distance modulus is
defined as \be \mu_{th}(z_i)\equiv m_{th}(z_i) - M =5 log_{10}
(D_L (z)) +\mu_0\,, \label{mth} \ee where \be \mu_0= 42.38 - 5
log_{10}h\,, \label{mu0}\ee  The steps followed for the usual
minimization of (\ref{chi2}) in terms of its parameters are
described in detail in \cite{Sanchez:2009ka,Lazkoz:2007cc,Nesseris:2004wj,Nesseris:2005ur}.

In the context of constraints from standard rulers from CMB spectrum peaks \cite{Lazkoz:2007cc,Sanchez:2009ka}, the maximum likelihood method is used and the
datapoints $(R,l_a,100\obh)$ of Ref. \cite{cmb} (WMAP5)
where $R$, $l_a$ are two shift parameters \cite{Lazkoz:2007cc}. In the case of BAO, the maximum likelihood method is also applied
\cite{Lazkoz:2007cc} using the datapoints of  Ref.
\cite{Percival:2007yw} (SDSS5). For comparison, the more recent data of Ref. \cite{bao}
(SDSS7) have also been considered with only minor differences in the results (slightly
reduced consistency with \lcdm in the context of the CPL
parametrization but no change in the ranking sequences).

\begin{figure}[h]
\includegraphics[width=25pc]{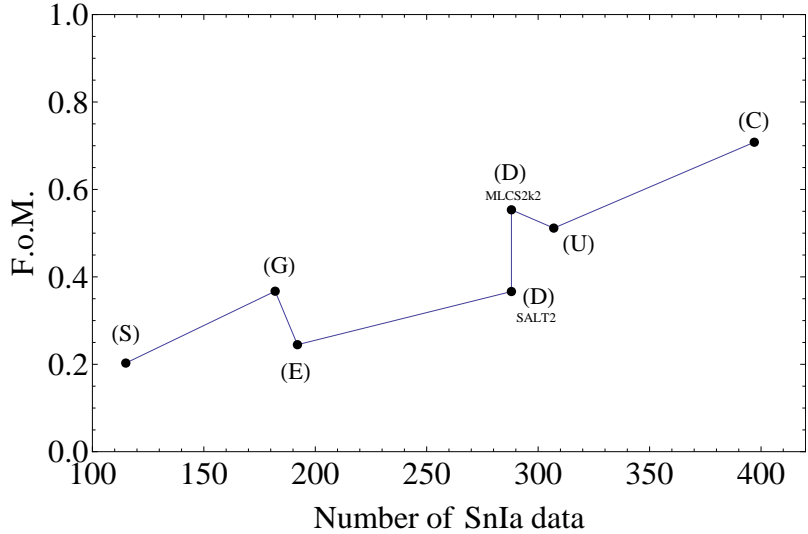}\hspace{2pc}%
\caption{The Figure of Merit (FoM) in terms of the number of the
SnIa data for $\omm=0.28$ using the CPL
parametrization.\label{fom} }
\end{figure}

\begin{figure*}[!t]
\rotatebox{0}{\hspace{0cm}\resizebox{0.45\textwidth}{!}{\includegraphics{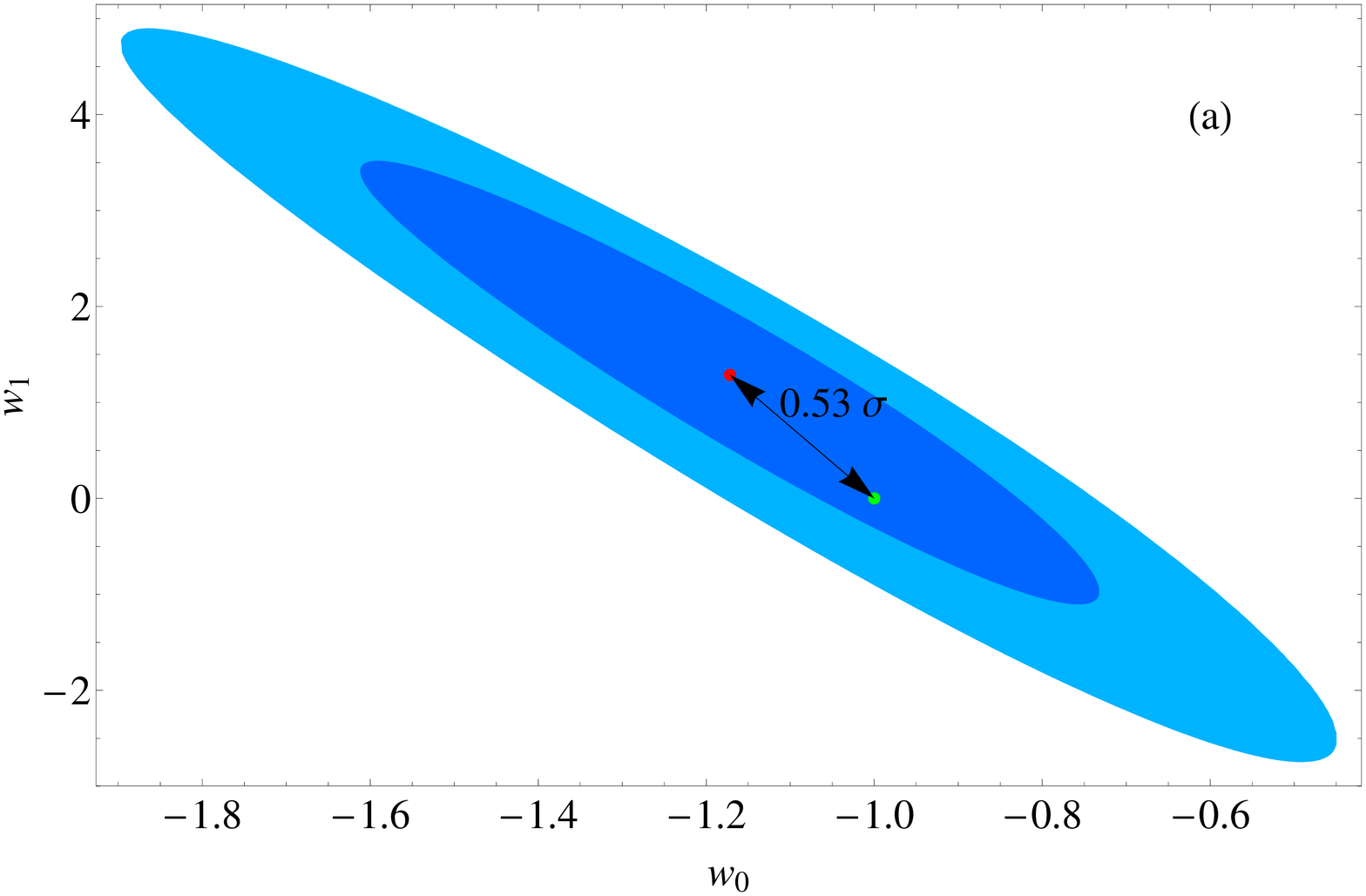}}}
\rotatebox{0}{\hspace{1cm}\resizebox{0.45\textwidth}{!}{\includegraphics{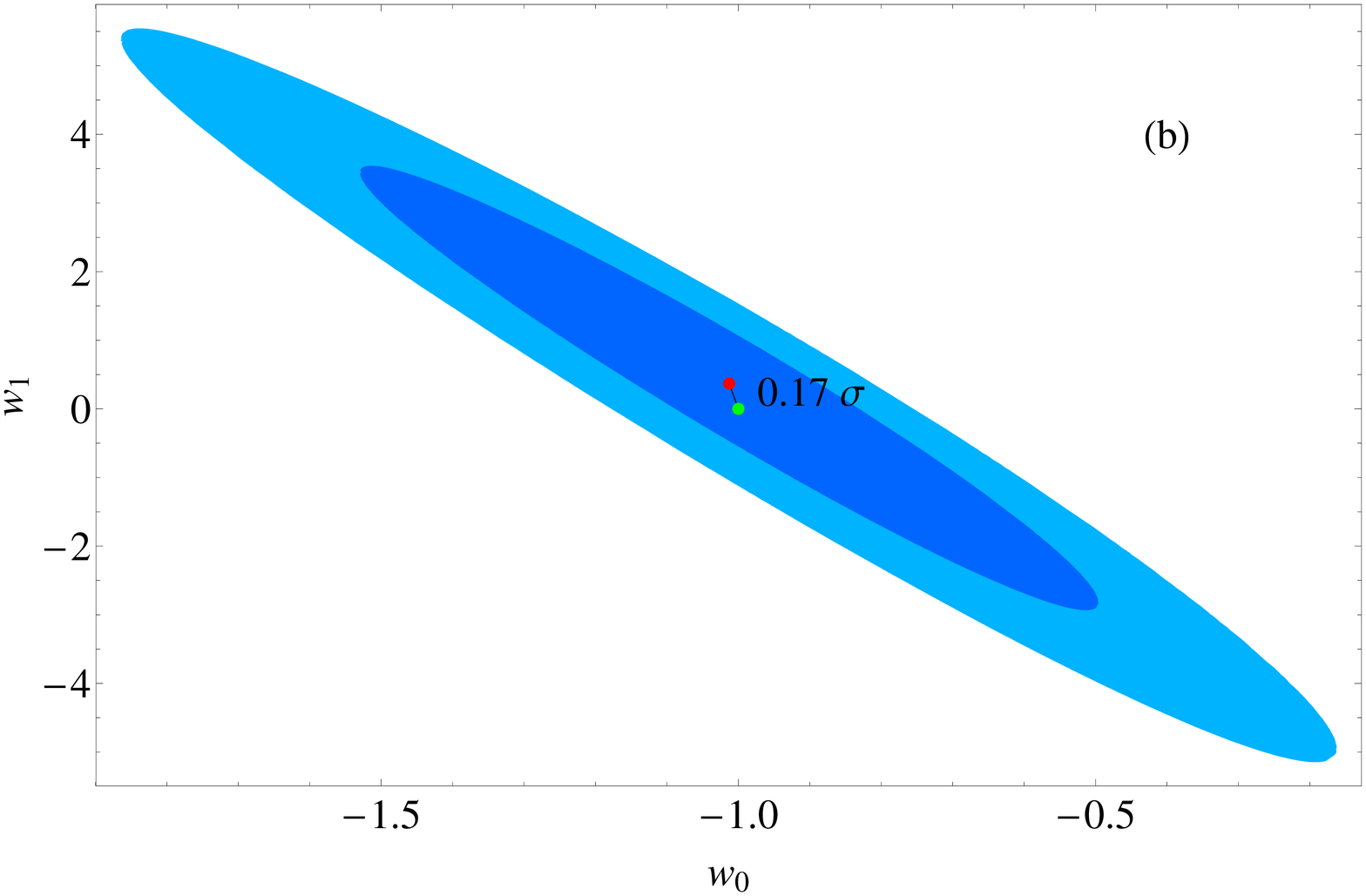}}}
\vspace{0pt}\caption{The $68.3\% (1\sigma)-95.4\%$ ($2\sigma$) $\chi^2$ confidence
contours in the $w_{0}-w_{1}$ plane based on parametrization
(\ref{hcpl}) for the ESSENCE (left) and SNLS1 datasets (right) for
$\omm=0.24$. The arrows indicate the $\sigma$-distance of $\Lambda$CDM
(green points: $(w_0,w_1)=(-1,0)$) to the best fit points (red
points). }\label{sdist}
\end{figure*}

The Figure of Merit (FoM) is a useful measure of the effectiveness
of a set of data in constraining cosmological parameters. In the
case of two parameters (as for the CPL parametrization) it is
defined as the reciprocal area of the $95.4\%$ contour (see Fig. \ref{sdist}), in
parameter space $(w_0,w_1)$ \cite{tfor}.
Clearly, the larger the FoM the more effective the dataset in
constraining the parameters $(w_0,w_1)$. Fig. \ref{fom} shows the FoM in terms of the
number of the SnIa data for the datasets of Table \ref{tabsnsets}. Clearly, the FoM is
an increasing function of the number of SnIa in the datasets. An
exception to this rule is the ESSENCE dataset which has a slightly
smaller FoM compared to the Gold06 dataset even though it has a
larger number of SnIa. A possible origin of this effect is the fact that the FoM does not depend only on the total
number of SnIa of the dataset but mainly on the number of SnIa at
low and high redshifts (the redshift
space distribution of the ESSENCE data includes more data at
intermediate redshifts than the Gold06 dataset).

In order to study the consistency of the various SnIa datasets
with the cosmological constant and the standard rulers we consider
the distance in units of $\sigma$ ($\sigma$-distance $d_\sigma$)
of the best fit point to a model with parameters $(w_0,w_1)$,
where this reference point can be either $\Lambda$CDM or some
other reference point, (see Fig.~\ref{sdist}). The
$\sigma$-distance can be found by converting $\Delta
\chi^2=\chi^2_{(w_0,w_1)}-\chi^2_{min}$ to $d_\sigma$, i.e.
solving \cite{press92} \be 1-\Gamma(1,\Delta
\chi^2/2)/\Gamma(1)={\rm Erf}(d_\sigma/\sqrt{2}) \label{sigmas}\ee
for $d_\sigma$ ($\sigma$-distance), where $\Delta \chi^2$ is the
$\chi^2$ difference between the best-fit and the reference point
($w_0,w_1)$ (eg $\Lambda$CDM) and ${\rm Erf}()$ is the error
function. The right hand side of Eq.~(\ref{sigmas}) comes from
integrating $\int_{-n \sigma}^{~n \sigma}\frac{1}{\sigma \sqrt{2
\pi}}~e^{-\frac{x^2}{2 \sigma^2}}dx$, where $n$ is the desired
number of $\sigma$s, while the left hand side corresponds to the
Cumulative Distribution Function (CDF) of a $\chi^2$
distribution\cite{press92} with two degrees of freedom. Note that
Eq.~(\ref{sigmas}) is only valid for the two parameters
$(w_0,w_1)$ and should be generalized accordingly for more
parameters \cite{press92}. In the special case of $n=1$ or $n=2$
we obtain the well known results $\Delta \chi_{1\sigma}=2.30$ and
$\Delta \chi_{2\sigma}=6.18$ valid for two parameter
parametrizations \cite{press92}.

Even though the $\sigma$-distance is not a commonly used statistic it
is quite useful because it can directly give information about
probability of a given region in parameter space. The integer
values of sigma distance ($1\sigma$ and $2\sigma$) are commonly
used to draw the corresponding contours in parameter space. The extension of this statistic to non-integer values is used to
find the specific contours that go through particular reference
points of parameter space and thus estimate quantitatively the
consistency of these points. The advantage of using the
$\sigma$-distance instead of $\Delta\chi^2$ is the fact that the
$\sigma$-distance takes into account the number of parameters of
the parameterizations and can therefore be directly translated
into probability for each point in parameter space. This is not
possible for $\Delta\chi^2$ because it does not include
information about the number of parameters of the
parameterizations considered.
\begin{figure}[!t]
\vspace{-0.2cm}\includegraphics[width=0.8\textwidth]{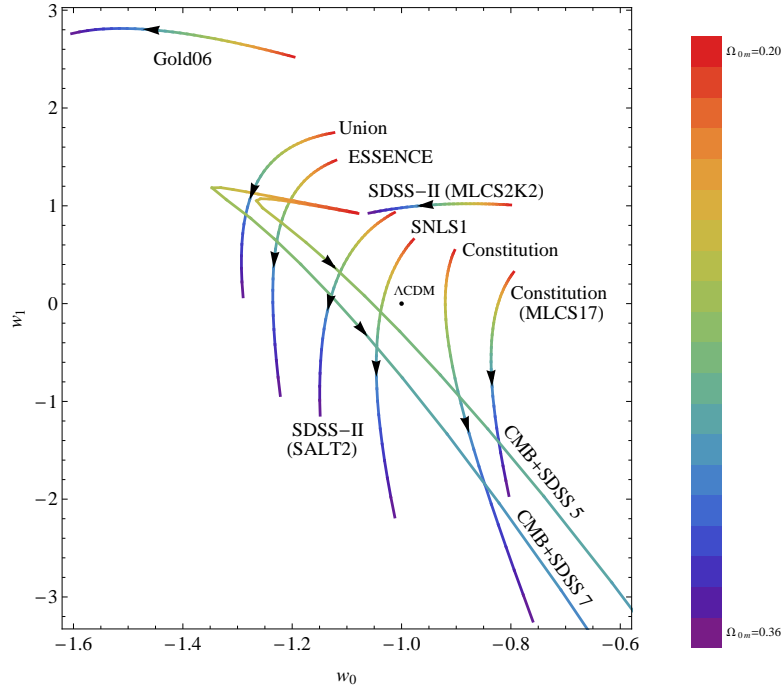}
\caption{Trajectories of the best fit points $(w_0,w_1)$ obtained
for each of the datasets of Table \ref{tabsnsets} and also for the
standard ruler CMB-BAO (WMAP5+SDSS5 and WMAP5+SDSS7) data as
$\omm$ varies in the range $\omm \in [0.2,0.36]$. The arrows in
the best fit lines indicate the direction of growing $\omm$. Note
that for the SDSS5 data the standard ruler best fit parameters
stretch out to $(w_0,w_1)\simeq(2,-30)$ for $\omm\simeq 0.36$,
whereas for the SDSS7 data
$(w_0,w_1)\simeq(0.90,-20)$.}\label{trajbf}
\end{figure}

It is straightforward to apply the likelihood method  to find the trajectory of the best fit point
$(w_0,w_1)$ in parameter space as $\omm$ varies in the range $\omm
\in [0.2,0.36]$. These trajectories obtained for each of the
datasets of Table \ref{tabsnsets} and also for the standard ruler
CMB-BAO (WMAP5+SDSS5, WMAP5+SDSS7) data are shown in
Fig.~\ref{trajbf}. These trajectories can not be used to directly
rank the datasets according to their consistency with any given
reference point in parameter space (e.g. $\Lambda$CDM) because
they contain no information about the $68.3\%$ ($1\sigma$) contours.
However, they provide useful hints for the trend of the best fit
parameters as $\omm$ varies. For example, such a trend is the
increase of the best fit value of the slope $w_1$ as the prior of
$\omm$ decreases towards the value $0.2$ or that the best fit
value of $w_0$ remains less than $-1$ for all datasets except of
the Constitution compilation.

\begin{figure*}[!t]
\rotatebox{0}{\hspace{0cm}\resizebox{0.45\textwidth}{!}{\includegraphics{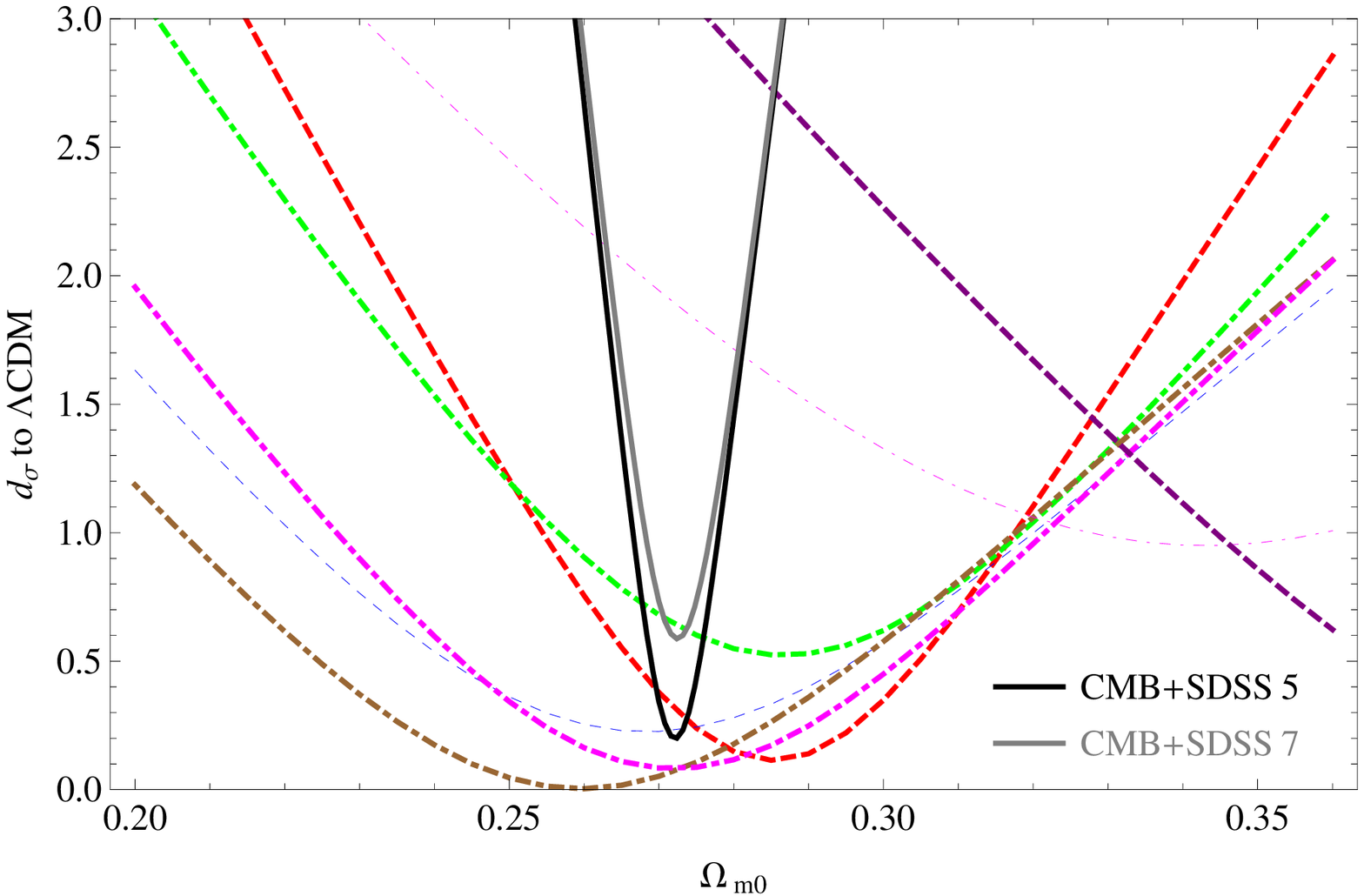}}}
\rotatebox{0}{\hspace{1cm}\resizebox{0.45\textwidth}{!}{\includegraphics{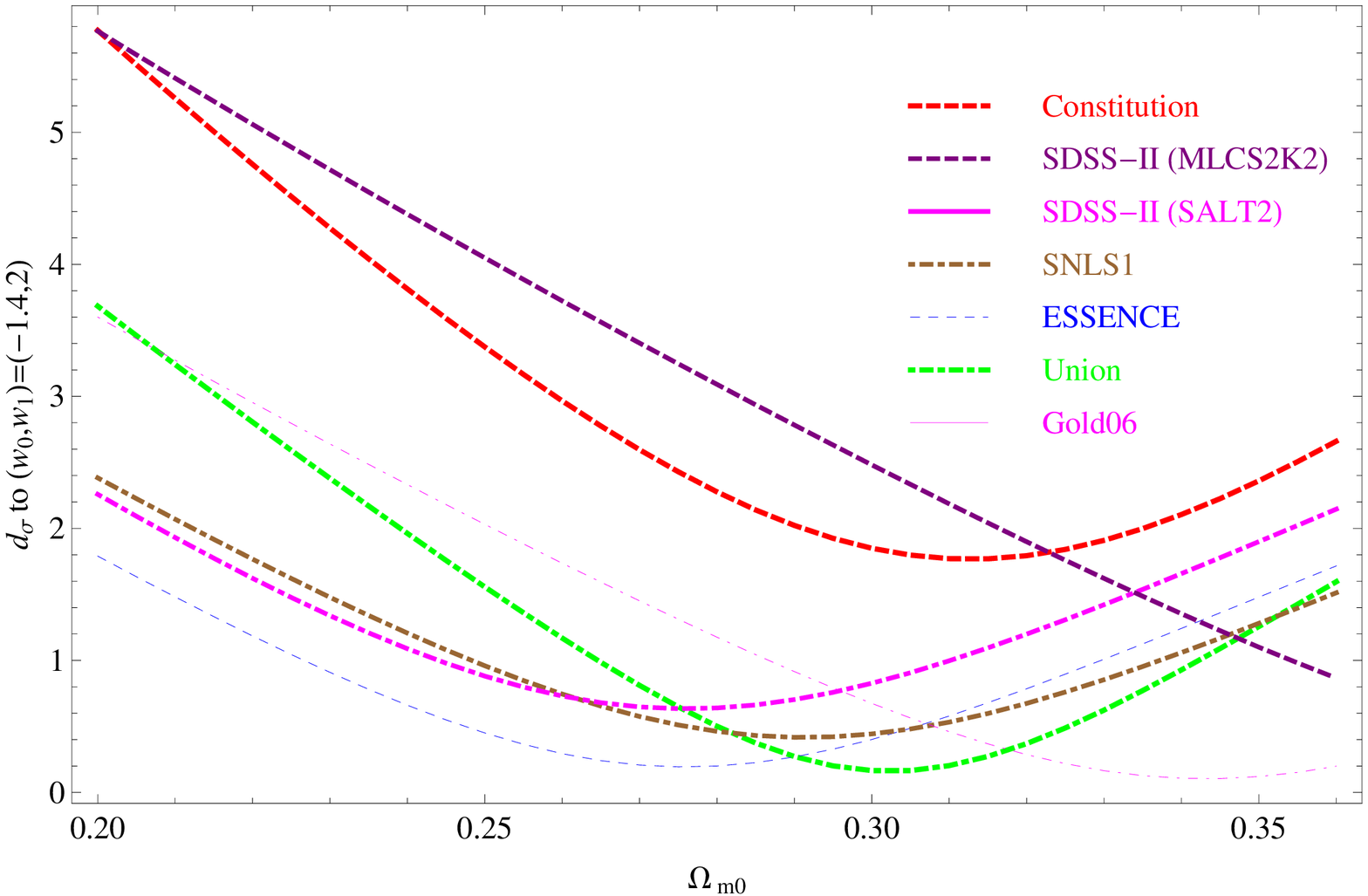}}}
\vspace{0pt}\caption{\small a: $\sigma$-distances $d_{\sigma
i}(\omm;-1,0)$ off the $\Lambda$CDM point (reference point) from
the best fit of each dataset in Table \ref{tabsnsets}. Notice that
the best fit of each dataset, except for the Gold06 and the
SDSS-II (MLCS2k2), are minimized at similar values of $\omm$. The
$\sigma$-distance $d_\sigma(\omm;-1,0)$ between the standard ruler
best fits and $\Lambda$CDM are also shown as a function of $\omm$
(black and grey solid lines).  b: Similar to (a) for the dynamical
dark energy reference point $(w_0, w_1)=(-1.4,2)$. Notice that the
$\sigma$-distances $d_{\sigma i}(\omm;-1.4,2)$ are minimized at
more widely separated values of $\omm$.} \label{dsmod}
\end{figure*}

The ranking sequence of the datasets of Table \ref{tabsnsets} with
respect to any reference point in parameter space can be studied
quantitatively using the $\sigma$-distance statistic discussed
above. In order to test the sensitivity of the ranking sequence of
datasets with respect to the choice of consistency reference point
$(w_0, w_1)$ we consider two such reference points: $(w_0,
w_1)=(-1,0)$ ($\Lambda$CDM) and $(w_0, w_1)=(-1.4,2)$ which
corresponds to dynamical dark energy with a $w(z)$ that crosses
the line $w=-1$. It should be noted that there is nothing special
about the parameter point (-1.4,2). We have selected it as a
representative of a wide region in parameter space (upper left
from $\Lambda$CDM) which corresponds to dynamical dark energy
crossing the phantom divide line $w=-1$. Any other point in the
same parameter region would lead to similar results and the same
ranking of datasets. This particular parameter region is
interesting because it is spanned by the best fit trajectories and
it also mildly favored by the Gold06 dataset (see Fig. \ref{trajbf}).

The resulting $d_\sigma(\omm)$ for each dataset of Table
\ref{tabsnsets} are shown in Figs.~\ref{dsmod} in
the range $\omm \in [0.2, 0.36]$.

\begin{table}[h]
\caption{Minimum $\sigma$-distance
$d_\sigma^{min}(\omm^{min};-1,0)$ from the best fit point for each
of the datasets to the $\Lambda$CDM point. Also listed are the
corresponding values of $\omm$, and the best fit parameters
$(w_0,w_1)$ (see also Fig.~\ref{dsmod}a). The SDSS-II (MLCS2K2)
data showed no minimum of $d_\sigma$ with respect to $\omm$ in the
range $\omm \in [0.2,0.36]$. We thus have simply displayed the
lowest value of $d_\sigma$ in the corresponding range of
$\omm$.\label{tabsep1}}
\begin{center}
\begin{tabular}{ccccc}
\br
\textbf{Dataset} \hspace{7pt}& $d_\sigma^{min}$ \hspace{7pt}& $\Omega_{0m}^{min}$ \hspace{7pt}& $w_0$
\hspace{7pt}& \hspace{7pt} $w_1$ \\
\mr
SNLS1 & 0.004 & 0.260 & -1.03 & 0.16 \\
SDSS-II (SALT2) & 0.084 & 0.270 & -1.09 & 0.51 \\
Constitution & 0.114 & 0.285 & -0.91 & -0.54 \\
ESSENCE & 0.227 & 0.270 & -1.20 & 1.04 \\
Union & 0.525 & 0.285 & -1.25 & 1.40 \\
SDSS-II (MLCS2K2) & 0.623 & 0.360 & -1.06 & 0.93 \\
Gold06 & 0.950 & 0.345 & -1.56 & 2.80 \\\hline\hline CMB+BAO
(SDSS5) & 0.200 & 0.272 & -1.15 & 0.51\\ CMB+BAO (SDSS7) & 0.588 &
0.272 & -1.30 & 0.97\\\hline
\br
\end{tabular}
\end{center}
\end{table}

Clearly, there are values of $\omm$ that minimize the
$\sigma$-distance $d_\sigma(\omm)$ between the best fit of each
dataset and the reference point. These values of $\omm$ maximize
the consistency of the datasets with the given reference point in
this range of $\omm$. The minima $\sigma$-distances
$d_\sigma(\omm)$ for each dataset, corresponding to maximum
consistency with $\Lambda$CDM along with the corresponding values
of $\omm$ are shown (properly ranked) in Table \ref{tabsep1}. The
corresponding results for the reference point $(w_0,
w_1)=(-1.4,2)$ are shown in Table \ref{tabsep2}.

\begin{table}[h]
\caption{The minimum $\sigma$-distances $d_\sigma^{min}$ to the
reference point $(w_0,w_1)=(-1.4,2)$, the values of $\omm$ at
which the minimum distance is attained, and the best fit
parameters $(w_0,w_1)$ at $\omm^{min}$ are displayed for each
dataset (see also Fig.~\ref{dsmod}b). We omit the rows
corresponding to CMB+BAO data as the resulting $\sigma$-distance
is always $\gg1$ due to the dominance of the dark energy at early
times. The SDSS-II (MLCS2K2) data showed no minimum of $d_\sigma$
with respect to $\omm$ in the range $\omm \in [0.2,0.36]$. We thus
have simply displayed the lowest value of $d_\sigma$ in the
corresponding range of $\omm$.\label{tabsep2}}
\begin{center}
\begin{tabular}{ccccc}
\br
\textbf{Dataset} \hspace{7pt}& $d_\sigma^{min}$ \hspace{7pt}& $\Omega_{0m}^{min}$ \hspace{7pt}& $w_0$
\hspace{7pt}& \hspace{7pt} $w_1$ \\
\mr
Gold06 & 0.11 & 0.345 & -1.56 & 2.80 \\
Union & 0.17 & 0.300 & -1.26 & 1.25 \\
ESSENCE & 0.19 & 0.275 & -1.21 & 0.99 \\
SNLS1 & 0.42 & 0.290 & -1.04 & -0.26 \\
SDSS-II (SALT2) & 0.63 & 0.275 & -1.10 & 0.46 \\
SDSS-II (MLCS2K2) & 0.87 & 0.360 & -1.06 & 0.93 \\
Constitution & 1.77 & 0.315 & -0.88 & -1.32\\\hline
\br
\end{tabular}
\end{center}
\end{table}

The following comments can be made with respect to the results
shown in Figs.~\ref{dsmod} and in the corresponding
Tables \ref{tabsep1}, \ref{tabsep2}.
\begin{enumerate}
\item The consistency with $\Lambda$CDM of all datasets, except
Gold06  and SDSS-II when using the MLCS2K2 method, is maximized in
a narrow range of $\omm \in [0.26, 0.29]$ which also includes the
value of $\omm$ favored by standard rulers. On the other hand, the
consistency with the dynamical dark energy point $(w_0,
w_1)=(-1.4,2)$ is maximized over a wider range of $\omm$ ($\omm
\in [0.27, 0.35]$) thus decreasing the consistency among the
datasets in the context of dynamical dark energy. \item The
ranking sequence changes dramatically when the consistency with
the dynamical dark energy is considered as a reference point
instead of $\Lambda$CDM (Table \ref{tabsep2}). Essentially the
ranking is reversed! Thus, the choice of the consistency reference
point plays an important role in determining the ranking sequence
of the datasets (see also Fig. \ref{dsmod}). \item The SDSS-II dataset
obtained with the MLCS2k2 fitter has some peculiar features
compared to other datasets. In particular it favors particularly
high values of $\omm$ ($\omm \simeq 0.4$) while for $\omm <0.3$
its consistency with \lcdm is significantly reduced to a level of
$3\sigma$ or larger ($d_\sigma > 3$). In addition, the trajectory
of its best fit parameter point as $\omm$ varies is perpendicular
to the corresponding trajectory of most other datasets (see Fig.
\ref{dsmod}).
\end{enumerate}

\begin{figure}[!t]
\vspace{-0.2cm}\includegraphics[width=0.8\textwidth]{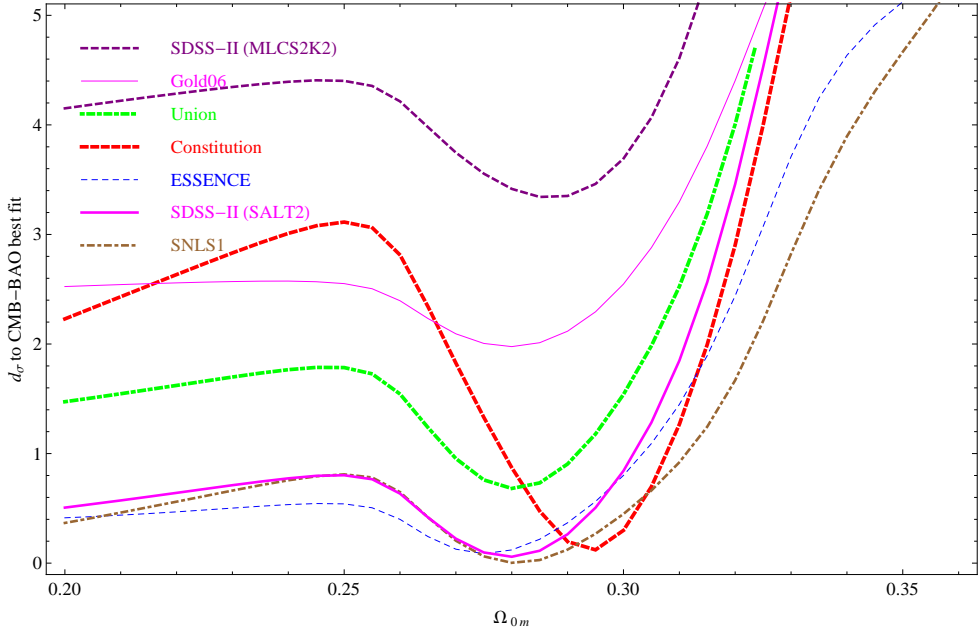}
\caption{$\sigma$-distance
$d_{\sigma i} (\omm; (w_0, w_1)^{SR})$ between the reference point
$(w_0,w_1)^{SR}$ of standard rulers (using the WMAP5+SDSS5 data)
and the best fit of each dataset as a function of $\omm$ for the
datasets of Table \ref{tabsnsets}. Using the SDSS7 data, the
minimum distances are found in the range $\omm\in[0.29,0.31]$.}
\label{dssr}
\end{figure}

It is straightforward to apply the $\sigma$-distance statistic to
rank the SnIa datasets according to their consistency with
standard ruler CMB-BAO data. We simply use as a consistency
reference point the best fit point $(w_0,w_1)^{SR}$ for standard
rulers obtained as described in section 2 using the WMAP5+SDSS5
data. In this case, the location of the reference point
$(w_0,w_1)^{SR}$ in parameter space depends on $\omm$ but this
does not complicate the analysis. The $\sigma$-distance between
the reference point $(w_0,w_1)^{SR}$ and the best fit of each
dataset is shown in Fig.~\ref{dssr} as a function of $\omm$ for
the datasets of Table \ref{tabsnsets}. These distances are
minimized for values of $\omm$ that are different for each dataset
but they are all in the narrow range $\omm \in [0.27,0.3]$.

These minimum distances along with the corresponding value of
$\omm$ are shown in Table \ref{tabsep3} for each dataset (properly
ranked according to consistency with standard rulers). Notice that
the ranking sequence for the consistency with standard rulers is
practically identical to the ranking sequence of the consistency
with $\Lambda$CDM (Table \ref{tabsep1}) but differs from the
ranking sequence of the consistency with dynamical dark energy
(Table \ref{tabsep2}). This is an interesting feature of the data
in favor of $\Lambda$CDM.

\begin{table}[h]
\caption{Consistency with standard rulers. Minimum
$\sigma$-distance $d_{\sigma i}^{min}(\omm;(w_0,w_1)^{SR})$
between best fit parameters for each dataset, $(w_0,w_1)$, and
best fit for standard rulers, $(w_0,w_1)^{SR}$. $d_{\sigma
i}^{min}(\omm;(w_0,w_1)^{SR})$ is minimized at $\omm^{min}$ (see
Fig. \ref{dssr}). \label{tabsep3}}
\begin{center}
\begin{tabular}{ccccccc}
\br
\textbf{Dataset} \hspace{7pt}& $d_\sigma^{min}$ \hspace{7pt}& $\Omega_{0m}^{min}$ \hspace{7pt}& $w_0$
\hspace{7pt}& \hspace{7pt} $w_1$ & \hspace{7pt}
$w_0^{SR}$ & \hspace{7pt} $w_1^{SR}$ \\
\mr
SNLS1 & 0.003 & 0.280 & -1.04 & -0.10 & -1.07 & 0.08 \\
SDSS-II (SALT2) & 0.058 & 0.280 & -1.11 & 0.40 & -1.07 & 0.08 \\
ESSENCE & 0.087 & 0.275 & -1.21 & 0.99 & -1.12 & 0.38 \\
Constitution & 0.121 & 0.295 & -0.90 & -0.76 & -0.84 & -1.28 \\
Union & 0.681 & 0.280 & -1.24 & 1.44 & -1.07 & 0.08 \\
Gold06 & 1.976 & 0.280 & -1.38 & 2.75 & -1.07 & 0.08\\
{\scriptsize SDSS-II(MLCS2K2)} & 3.342 & 0.285 & -0.92 & 1.02 &
-1.00 & -0.28\\\hline
\br
\end{tabular}
\end{center}
\end{table}
It is therefore clear that the \lcdm cosmological model is a well defined, simple and predictive model which is consistent with the majority of current cosmological observations. Despite of these successes there are specific cosmological observations which differ from the predictions of \lcdm at a level of $2\sigma$ or higher.
The observations conflicting the WMAP5 normalized \lcdm model at a level of $2\sigma$ or larger include the following \cite{Perivolaropoulos:2008ud}:
\begin{itemize}
\item Large Scale Velocity Flows (\lcdm predicts significantly smaller amplitude and scale of flows than what observations indicate)\cite{Watkins:2008hf,Kashlinsky:2008ut,Feldman:2009es}. The probability of consistency with \lcdm is about $1\%$. \item Brightness of Type Ia Supernovae (SnIa) at High Redshift $z$ (\lcdm predicts fainter SnIa at High $z$)\cite{Perivolaropoulos:2008yc}. The probability of consistency with \lcdm is about $3-5\%$ for the Union and Gold06 datasets. \item Emptiness of Voids (\lcdm predicts more dwarf or irregular galaxies in voids than observed)\cite{Tikhonov:2008ss,Peebles:2003pk,Klypin:1999uc}. \item Profiles of Cluster Haloes (\lcdm predicts shallow low concentration and density profiles in contrast to observations which indicate denser high concentration cluster haloes) \cite{Broadhurst:2008re,Broadhurst:2004bi}. \item Profiles of Galaxy Haloes\cite{Gentile:2007sb} (\lcdm predicts halo mass profiles with cuspy cores and low outer density while lensing and dynamical observations indicate  a central core of constant density and a flattish high dark mass density outer profile),\item Sizable Population of Disk Galaxies\cite{Bullock:2008wv} (\lcdm predicts a smaller fraction of disk galaxies due to recent mergers expected to disrupt cold rotationally supported disks).\end{itemize}
     Even though some of the puzzles discussed here may be resolved by more complete observations or astrophysical effects, the possible requirement of more fundamental modifications of the \lcdm model remains valid.

\section{Dynamical Probes: The Growth Function}
\subsection{Growth of Perturbations in General Relativity: Beyond the sub-Hubble Approximation}
As discussed in the introduction, the evolution of the matter overdensity $\delta_m(k,a)$ (the growth function) consists a useful probe of both the expansion rate and the gravitational law on large scales.
 The standard parametrization of the linear growth function $\delta_m (k,a)$ is scale independent and is obtained by introducing a growth index $\gamma$ defined through the growth rate $f(a)$ by  \be f_{0}(a)\equiv \frac{d\ln \delta_{m}}{d\ln a}=\omms(a)^\gamma \label{f0def}\ee where $a=\frac{1}{1+z}$ is the scale factor and
\be \omms (a) \equiv \frac{H_0^2 \omm a^{-3}}{H(a)^2} \label{omadef} \ee
 is the ratio of the matter density to the critical density when the universe has scale-factor $a$ where $H_0$ is Hubble constant and  $\Omega_{0m}$ is ratio of mass density to critical density.
This parametrization \cite{Wang:1998gt} provides an excellent fit to the evolution equation for $\delta_m(a)$ in general relativity and in the small scale (sub-Hubble) approximation \be {\ddot \delta}_m + 2 H {\dot \delta}_m - 4\pi G \rho_m \delta_m =0 \label{greqtim} \ee  where  an overdot denotes the derivative with respect to time and $\rho_m$ is the matter density. Changing variables from $t$ to $\ln a$ we obtain the evolution equation for the growth factor $f$ as
\be f' + f^2 + f(\frac{\dot H}{H^2}+2)=\frac{3}{2} \omms \label{greqlna} \ee  where $' =d/dlna$.
For dark energy models in a flat universe with a slowly varying equation of state $w(a)\equiv \frac{p(a)}{\rho(a)}=w_0$, the solution of eq. (\ref{greqlna}) is well approximated by  eq. (\ref{f0def}) with \cite{Wang:1998gt} \be \gamma=\frac{3(w_0 -1)}{6w_0-5} \label{gamval} \ee which reduces to $\gamma=\frac{6}{11}$ for the \lcdm case ($w_0=-1$). It is therefore clear that the observational determination of the growth
index $\gamma$ can be used to test \lcdm \cite{Nesseris:2007pa}. It has been shown \cite{lindercahn} that even in the context of dynamical dark energy models consistent with Type Ia supernovae (SnIa) observations the parameter $\gamma$ does not vary by more than $5\%$ from its \lcdm value. However, in the context of modified gravity models $\gamma$ can vary by as much as $30\%$ (e.g. for the DGP model\cite{dgp} $\gamma_{DGP}\simeq 0.68$ \cite{lindercahn}) while scale dependence is also usually introduced\cite{Nesseris:2006er,Uzan:2006mf,Polarski:2007rr}.

Current observational constraints on $\gamma$ are based on redshift
distortions of galaxy power spectra \cite{Hawkins:2002sg}, the rms
mass fluctuation $\sigma_8(z)$ inferred from galaxy and
$Ly-\alpha$ surveys at various redshifts
\cite{Viel:2004bf,Viel:2005ha}, weak lensing statistics
\cite{Kaiser:1996tp}, baryon acoustic oscillations
\cite{Seo:2003pu}, X-ray luminous galaxy clusters
\cite{Mantz:2007qh}, Integrated Sachs-Wolfe (ISW) effect
\cite{Pogosian:2005ez} etc. Unfortunately, the currently available
data are limited in number and accuracy and come mainly from the
first two categories. They involve significant error bars and
non-trivial assumptions that hinder a reliable determination of
$\gamma$. Thus, the current constraints on $\gamma$ are fairly weak \cite{Nesseris:2007pa} and are expressed as \be \gamma=0.674^{+0.195}_{-0.169}
\label{gambf}\ee This however is expected to change in the next few years when more detailed weak lensing surveys \cite{weak-lens} are anticipated to narrow significantly the above range\cite{futde}.

A crucial assumption made in the derivation of eq. (\ref{greqtim}) in the context of general relativistic metric perturbations is the assumption that the scale of the perturbations is significantly smaller than the Hubble scale \cite{Wang:1998gt}. This assumption however does not lead to a good approximation on scales larger than about $100h^{-1}Mpc$ \cite{Dent:2008ia,Dent:2009wi}. In order to demonstrate this fact consider the perturbed metric of spacetime which takes the form (in the Newtonian gauge):
\begin{equation}
 ds^2 = -(1+2\Phi) dt^2 + (1-2\Phi)a^2\gamma_{ij}dx^i dx^j,
\end{equation}
where $\gamma_{ij}$ is the metric of the spatial section {and we are ignoring anisotropic stresses}. The evolution of density perturbations
on all scales is dictated by combining the background equations \ba H^2 &=& \frac{8\pi G}{3}(\rho_m +\rho_{de}) \label{bcgeq1} \\ {\dot \rho} &=& -3H (\rho + p) \label{bcgeq2} \ea    (assuming a flat universe with only
pressureless dark matter  and (non-clustering) dark energy) with the perturbed  linear order Einstein equations in the Newtonian gauge \cite{mabertschinger} ($\rho_m$ and $\rho_{de}$ are the matter and dark energy densities respectively while $p=w \rho$ is the pressure). The resulting (anisotropic stress-free) equations are of the form
\ba
\label{grper1}\ddot{\Phi}&=&-4H\dot{\Phi}+8\pi G \rho_{de} w_{de}\Phi\\
\label{grper2}\dot{\delta}&=&3\dot{\Phi}+\frac{k^2}{a^2}v_{f}\\
\label{grper3}\dot{v}_{f}&=&-\Phi
\ea
with constraint equations
\ba
\label{grcons1}3H(H\Phi+\dot{\Phi})+\frac{k^2}{a^2}\Phi&=&-4\pi G\delta\rho_m\\
\label{grcons2}(H\Phi+\dot{\Phi})&=&-4\pi G\rho_m v_{f}
\ea
where $\Phi$ is the Newtonian potential, $v_f\equiv-v a$ ($v$ is the velocity potential for dark matter). Clearly, equations (\ref{grper1})-(\ref{grper3}) involve a scale $k$ dependence in contrast to the small scale approximate equation (\ref{greqtim}) which is scale independent.

The derivation of equation (\ref{greqtim}) in the context of general relativity is made using the  sub-Hubble approximation. The linear matter overdensity $\delta \rho_m$ may be expressed \cite{mabertschinger} in terms of the gravitational potential $\Phi$ and the background variables as follows: \be
-4\pi G\delta\rho_m = \frac{k^2}{a^2}\Phi +3H^2\Phi +3H\dot{\Phi} \label{drhom}
\ee
In the sub-Hubble (small scale) approximation ($\frac{k^2}{a^2} \gg H^2$) equation (\ref{drhom}) takes the form
\be
-4\pi G\delta\rho_m = \frac{k^2}{a^2}\Phi \label{drhomss}
\ee where a slowly varying gravitational potential $\Phi$ has also been assumed.

The general relativistic equations (\ref{grper1})-(\ref{grcons2})
lead to the following equation for the matter overdensity $\delta_m$ \be {\ddot \delta_m} + 2 H {\dot \delta_m} + \frac{k^2}{a^2} \Phi = 0 \label{grdel1} \ee
which also expresses the conservation of the perturbed energy momentum tensor for matter.
Using the sub-Hubble approximation (\ref{drhomss}) we obtain the scale independent approximate equation (\ref{greqtim}). On the other hand, if we avoid this approximation in equation (\ref{drhom}), solve for $\Phi$ (ignoring the time derivative) and substitute in equation (\ref{grdel1}) we obtain the following scale dependent evolution equation for $\delta_m$ \cite{Dent:2008ia,Dent:2009wi}:
\be {\ddot \delta_m} + 2 H {\dot \delta_m} - \frac{4\pi G \rho_m \delta_m}{1+\xi(a,k)} =0 \label{greq-scdep} \ee
where \be \xi(a,k)=\frac{3 a^2 H(a)^2}{k^2} \label{xidef} \ee
The solution of equation (\ref{greq-scdep}) provides a much better approximation to the full linear general relativistic system (\ref{grper1})-(\ref{grper3}) up to horizon scales.  On scales larger than the horizon even equation (\ref{greq-scdep}) breaks down since on these scales, the time derivative of $\Phi$ can not be ignored.

Given the successful approximation of the solution of (\ref{greq-scdep}) to the exact linear general relativistic solution, it becomes important to construct a scale dependent parametrization that is analogous to (\ref{f0def}) and solves (approximately) (\ref{greq-scdep}) for all scales $k$. In order to construct such a parametrization we focus on the matter dominated era when most of the growth occurs and express $\xi(a,k)$ as
\be \xi(a,k)=\frac{3 H_0^2 \omm}{a k^2} \label{xidef2} \ee Equation (\ref{greq-scdep}) may be expressed in terms of the growth factor $f=\frac{d\ln \delta_m}{d\ln a}$ in the form
\be f' + f^2 + \(2-\frac{3}{2} \omms(a)\)f=\frac{3}{2}\frac{\omms(a)}{1+\xi(a,k)} \label{greqlna-scdep} \ee
where $'\equiv \frac{d}{d\ln a}$, and we have assumed \lcdm for $H(a)$.

For sub-Hubble scales $\xi(k,a)\rightarrow 0$ and equation (\ref{greqlna-scdep}) reduces to (\ref{greqlna}) whose solution is well approximated by (\ref{f0def}) with $\gamma=\frac{6}{11}$. It is straightforward to show using an expansion method \cite{Dent:2009wi} that the parametrization \be f(k,a)=\frac{f_0(a)}{1+\xi(k,a)}=\frac{\omms(a)^\gamma}{1+\frac{3 H_0^2 \omm}{a k^2}} \label{newpar} \ee is an approximate solution of equation (\ref{greqlna-scdep}) and provides a good approximation to the solution of the general relativistic system (\ref{grper1})-(\ref{grper3}) up to horizon scales.

\begin{figure*}[!t]
\rotatebox{0}{\hspace{0cm}\resizebox{0.35\textwidth}{!}{\includegraphics{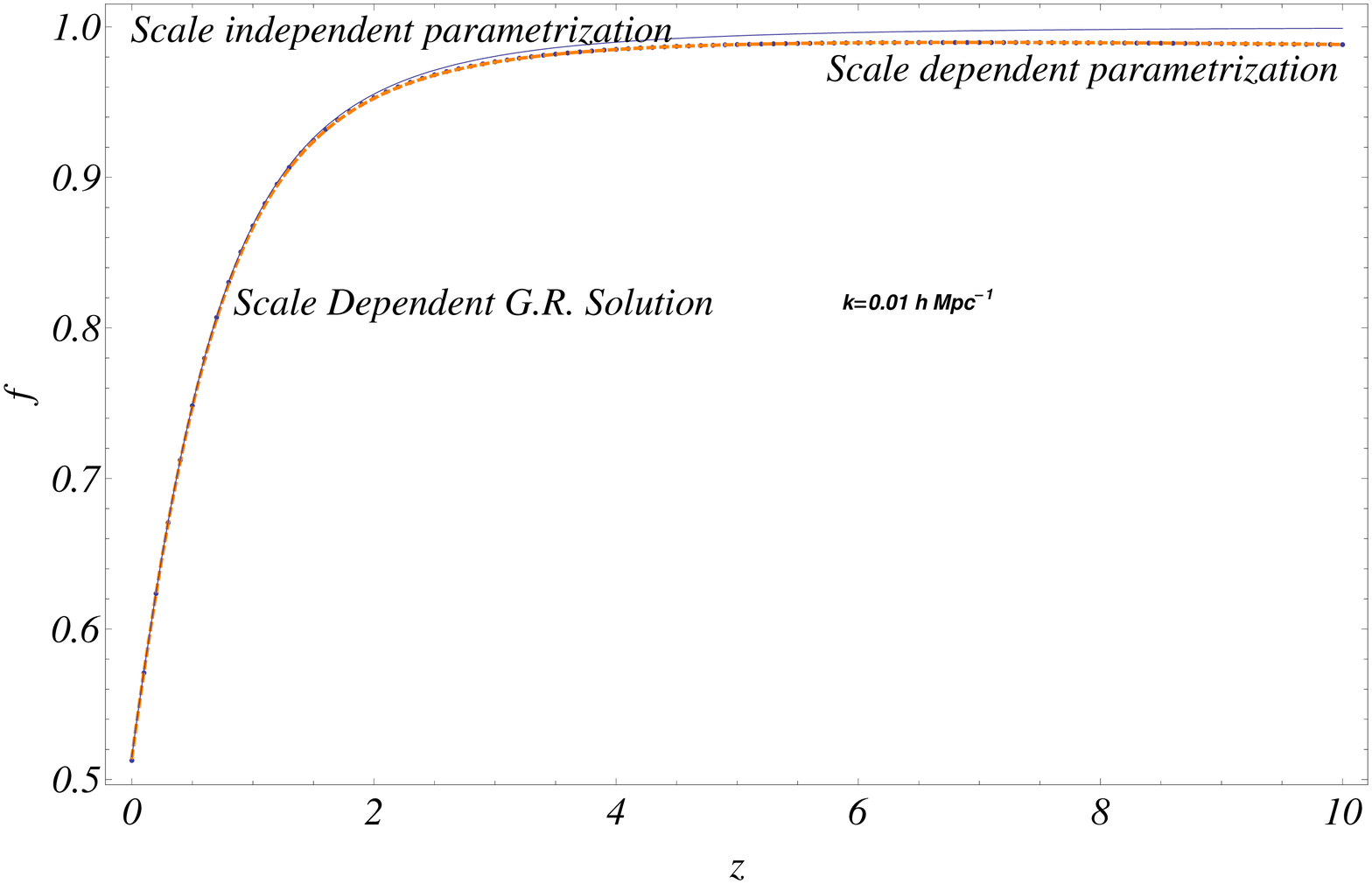}}}
\rotatebox{0}{\hspace{0cm}\resizebox{0.35\textwidth}{!}{\includegraphics{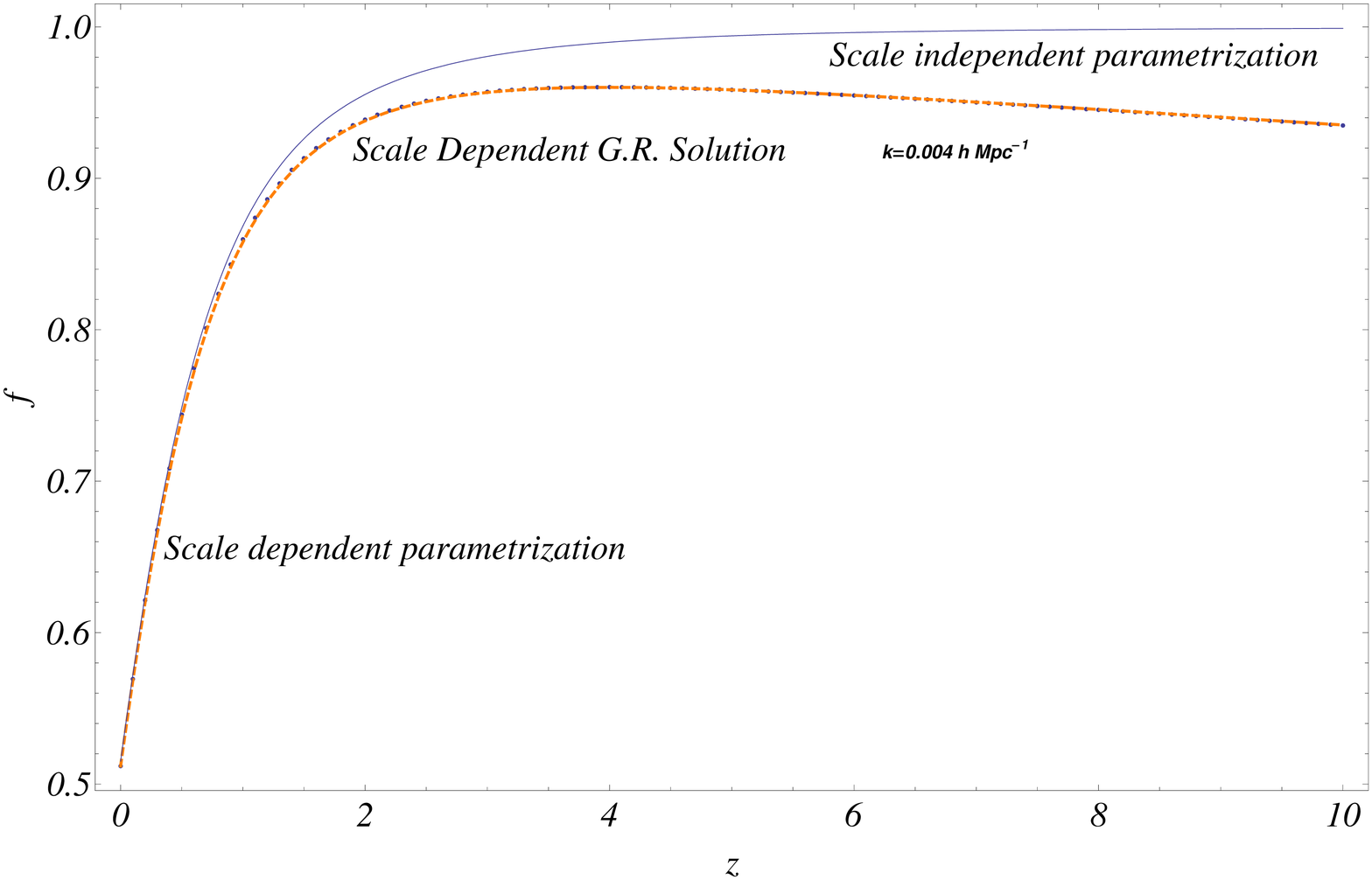}}}
\rotatebox{0}{\hspace{0cm}\resizebox{0.35\textwidth}{!}{\includegraphics{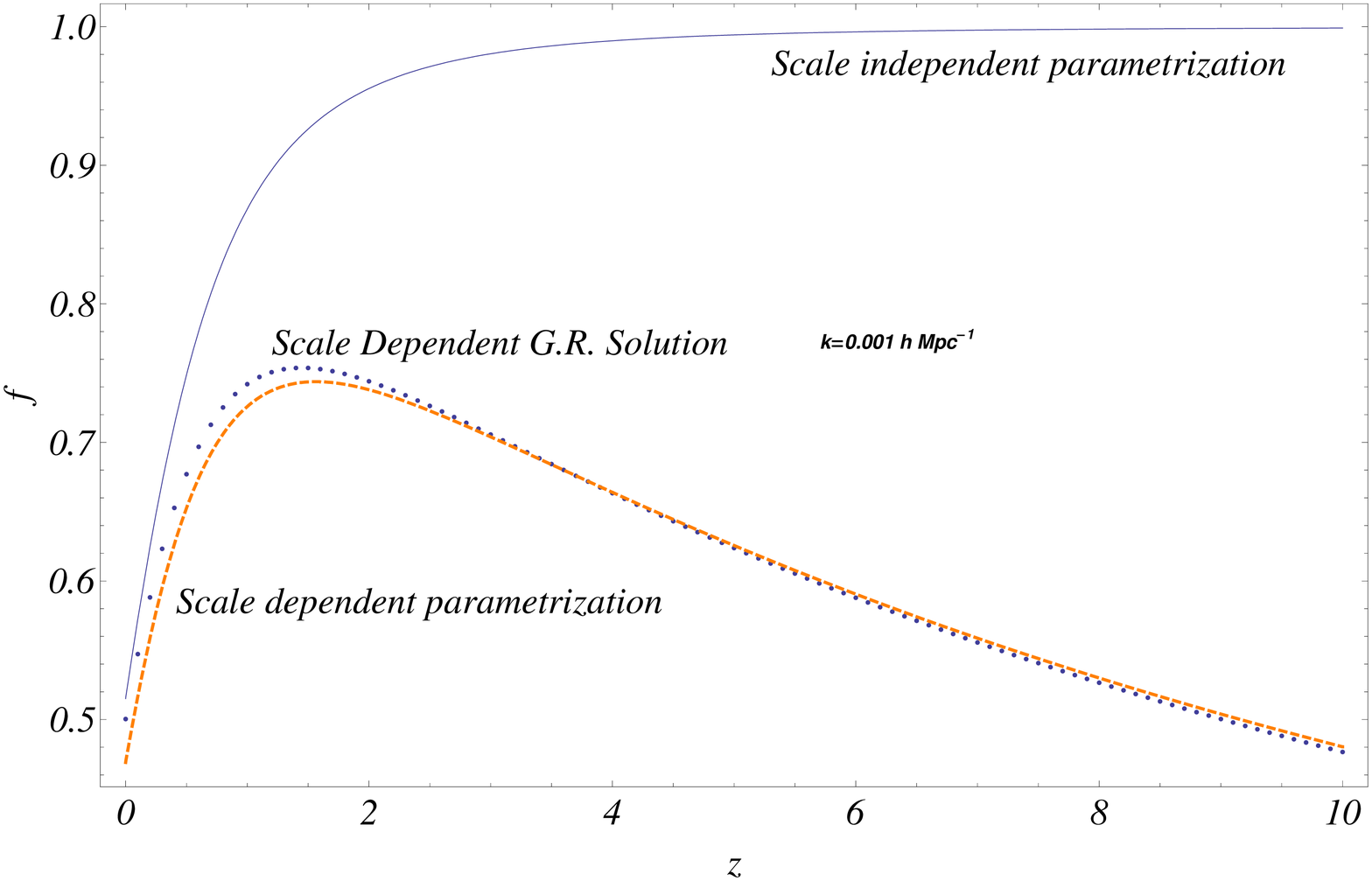}}}
\vspace{0pt}\caption{a: The growth rate $f$ obtained from the solution of the general relativistic system ($k=0.01h^{-1}Mpc$, $\omm=0.3$, \lcdm, dotted line) compared with the scale independent parametrization (continuous line) and the corresponding generalized scale dependent parametrization (thick dashed line) b: Similar to a. for $k=0.004h^{-1}Mpc$. c: Similar to a. for $k=0.001h^{-1}Mpc$.  }\label{newparfig}
\end{figure*}

The accuracy of the parameterization (\ref{newpar}) is demonstrated in Fig. \ref{newparfig} where we compare the form of the scale dependent parametrization \eqn{newpar} with the general relativistic numerical solution and with the standard parametrization for three different scales. It is clear that up to approximately the Hubble scale $(k \simeq 0.001hMpc^{-1})$ the scale dependent parametrization (\ref{newpar}) is accurate at a level better than $5\%$ at least up a to redshift $z=10$. It can be shown that this parametrization is also a good approximation to the growth rate not only in the $\Lambda$CDM model but also in the case of dynamically evolving dark energy. Thus the parametrization described by equation (\ref{newpar}) approximates very well the full linear general relativistic solution up to horizon scales and differs significantly from the standard parametrization on scales larger than about $100 h^{-1}Mpc$. Its use may play an important role when comparing cosmological structure data on large scales with theoretical model predictions.

As shown in Fig. \ref{newparfig} the range of scales where the standard scale independent parametrization (\ref{f0def}) starts breaking down involves scales larger than $100 h^{-1} Mpc$ ($k<0.01 h Mpc^{-1}$). This scale is much smaller than the present Hubble scale which is about $3000 h^{-1} Mpc$. Therefore, the sub-Hubble approximation breaks down at much smaller scales than the anticipated Hubble scale. This is due to the fact that at early times during the matter era when most growth occurs, the comoving Hubble scale is significantly smaller compared to its present value. In fact, at recombination it is close to $200 h^{-1}Mpc$. Therefore, scales that remain at the sub-Hubble level during the whole time when the growth of matter perturbations take place are only scales below $200 h^{-1}Mpc$.

The generalized Poisson equation (\ref{drhom}) in the Newtonian gauge may be solved for the slowly varying $\Phi$ leading to a generalized gravitational potential in Fourier space of the form \be \Phi^{GR}(k) =\frac{\Phi^{Poisson}(k)}{1+\frac{3a^2H(a)^2}{k^2}} \label{generphik} \ee where $\Phi^{Poisson}(k)$ is the usual small scale gravitational potential which emerges as a solution of the usual Poisson equation. In coordinate space the generalized large scale potential potential has a Yukawa form with a Hubble scale cutoff namely \be \Phi^{GR}(r)=-\frac{GM}{a r}e^{-\sqrt{3}H a r} \label{generphir} \ee It is this general relativistic potential that should be compared with corresponding modified gravity potentials on scales comparable to the Hubble scale.

In a modified gravity theory the Poisson equation on large scales is modified for two reasons: First due to the modified gravitational law and second due to the Hubble scale effects that are present also in the case of general relativity as discussed above. The result is a Poisson equation of the form \be \frac{k^2}{a^2} \Phi = -4\pi G {\rho}_m \delta_m(k,a) f(k,a) \label{modgravpois} \ee where the scale dependent function $f(k,a)$ incorporates the scale dependence due to both Hubble scale effects and scale dependence due to modification of the gravitational law. This function is to be compared with the scale dependence due to pure Hubble scale effects $f_{GR}(k,a)=\frac{1}{1+\xi(a,k)}$ present in general relativity. Therefore a signature of modified gravity would not be a scale dependence of the evolution of matter density perturbations as claimed occasionally in the literature\cite{Acquaviva:2008qp}, but a deviation from the scale dependence predicted on large scale in the context of general relativity.

\subsection{Gauge Dependence}
The scale dependence of the matter density growth rate obtained above was based on calculations made in the Newtonian gauge. This is a physically important gauge because it corresponds to a time slicing of isotropic expansion. Nevertheless since the matter density perturbation $\delta_m(k,t)$ is a gauge dependent quantity, it is important to clarify to what extend do the results of section 3.1 persist in a gauge different from the Newtonian gauge.

Another important gauge is the synchronous gauge which corresponds to a time slicing obtained by the matter local rest frame everywhere in space (the free falling observer frame). The synchronous gauge is generally considered to be the most efficient reference system for doing calculations. It is used in many modern cosmology codes for calculating the evolution of cosmological perturbations (eg  CMBFAST \cite{Zaldarriaga:1999ep}). The line element of the perturbed spacetime in the synchronous gauge is given by
\begin{eqnarray}
ds^2 = a^2(\tau)(-d\tau^2 + (\delta_{ij} + h_{ij})dx^idx^j)
\end{eqnarray}
where $\tau$ is the conformal time.
It is straightforward to derive the growth equation for $\delta_m \equiv \delta\rho_m/\rho_m$ in a matter dominated universe in the synchronous gauge to obtain \cite{mabertschinger,Dent:2008ia} \be
\label{sgrowth}
\ddot{\delta}^{SG}_m + 2H\dot{\delta}^{SG}_m -4\pi \rho_m G \delta^{SG}_m = 0
\ee
This growth equation is exact in the synchronous gauge in the case of matter domination and involves no scale dependence as in the case of equation  (\ref{greq-scdep}) of the Newtonian gauge. This scale independence is an artifact of the particular time slicing of the synchronous gauge which is a good approximation on small scales but is unable to capture the horizon scale effects modifying the growth function on large scales.

Nevertheless equations (\ref{greq-scdep}) and (\ref{sgrowth}) clearly agree on small scales where $\xi\rightarrow 0$. Therefore, for larger scales ($k<0.01 h Mpc$) the question that arises is the following: {\it What is the proper gauge to use when comparing with observations?} This question was recently addressed in Ref. \cite{Yoo:2009au} where a gauge invariant observable replacement was obtained for $\delta_m$. This observable $\delta_{obs}(k,t)$ involves the matter density perturbation $\delta_m(k,t)$ corrected for redshift distortions due to peculiar velocities and gravitational potential. It also includes volume and position corrections. The final expression however is complicated and makes the theoretical predictions based on it not easy to implement and manipulate.

Alternatively, the gauge-invariant (GI) approach to the
cosmological perturbations evolution,
pioneered by Bardeen\cite{bardeen2} may be used to identify observables (e.g., Refs.\cite{kodama1,mukhanov}). The most general form of the line element for a spatially flat
background and scalar metric perturbations can be written as
\cite{mukhanov}
\begin{equation}
  ds^2=a^2\{(1+2\Phi)d\tau^2-2B_{\mid i}dx^id\tau-[\delta_{ij}
-2(\Psi\delta_{ij}-E_{\mid_{ij}})]dx^idx^j~\},
\end{equation}
where $a$ and $\tau$ are the conformal cosmic expansion scale factor
and the conformal cosmic time; ``$_\mid$'' denotes the background
three-dimensional covariant derivative. The corresponding perturbed
energy-momentum tensor $T^{\mu}_{\nu}$ has the form
\begin{eqnarray}
  \nonumber T^0_{~0}&=&\rho_m(1+\delta_m)~,\\
  \nonumber T_0^{~i}&=&\rho_m U_{\mid i}~,\\
  \nonumber T^0_{~i}&=&-\rho_m (U-B)_{\mid i}~,\\
  T^i_{~j}&=&-\rho_m
        \Sigma_{\mid ij}~,
\end{eqnarray}
where $\rho_m$ is the unperturbed pressureless matter density; $U$ and
$\Sigma$ determine velocity perturbation and anisotropic shear
perturbation.

A gauge-invariant matter density
perturbation
may be constructed as \cite{bardeen2,kodama1}
\begin{equation}\label{epsilonm}
  \delta_m^{GIS}\equiv \delta_m +3
        \frac{\dot{a}}{a}(U-B)~,
\end{equation}
$\delta_m^{GIS}$ coincides with the density perturbation $\delta_m^{(CTG)}$
in the Comoving Time-orthogonal Gauge ($CTG$, in which $U=B=0$), which
denotes the density perturbation relative to the spacelike hypersurface
which represents the
matter local rest frame everywhere\cite{bardeen2}. This quantity also
coincides with the density perturbation $\delta_m^{SG}$ in the
Synchronous Gauge (in which $\Psi=\Phi=B=0$) for the pressureless matter
system. In
other words, $\delta_m^{GIS}$ denotes the density perturbation relative to
the observers everywhere comoving with the matter. These {\it free falling} observers do not experience the
isotropic expanding background of the universe because the peculiar velocity of matter is distinct from the Hubble flow. Thus $\delta_m^{SGI}$ has
physical
significance only for perturbations on scales small compared to the Hubble scale.

An alternative gauge-invariant variable is more closely related to the matter overdensity in the Newtonian gauge.
This gauge-invariant perturbation variable is of the form
\cite{bardeen2,kodama1,mukhanov},
\begin{equation}\label{gidp}
  \delta_m^{GIN}\equiv
  \delta_m+\frac{\dot \rho}{\rho}(B-\dot{E})=
\delta_m-3\frac{\dot{a}}{a}(B-\dot{E})~.
\end{equation}
and has important advantages over $\delta_m^{GIS}$.
$\delta_m^{GIN}$ coincides with the density perturbation
$\delta_m^{NG}$ in the Newtonian Gauge
($NG$), in which $B=E=0$.

In addition to the gauge invariant perturbation variable it is straightforward to construct
two gauge-invariant scalar potentials $\phi$ and
$\psi$,
both of which become the same as the gravitational potential
in the Newtonian limit. These are constructed from metric perturbations as follows
\cite{mukhanov}:
\begin{eqnarray}\label{potential}
\nonumber && \phi \equiv \Phi-\frac{\dot{a}}{a}(B-\dot{E})~,\\
      && \psi \equiv \Psi+\frac{1}{a}\frac{d}{d\tau}[(B-\dot{E})a]~.
\end{eqnarray}

The relation between
$\delta_m^{GIS}$ and the general gravitational potential $\phi$
obeys
the Poisson equation\cite{bardeen2,kodama1}:
\begin{equation}\label{poisson}
\bigtriangledown^2\phi=-k^2\phi
=4\pi G\rho a^2\delta_m^{GIS}.
\end{equation}
where $k$ is the (comoving) wavenumber of Fourier mode. As discussed above, the Poisson
equation is valid only for scales small
compared to the Hubble radius $1/H$ while on scales larger than the Hubble scale the growth of matter density perturbations is frozen. Hence, $\delta_m^{GIS}$ can not be regarded as the observable matter density
perturbation on scales comparable to
the Hubble scale.  Therefore,
the
observable density perturbation on both the small-scale and
the large-scale modes can not be described {\em directly} by
$\delta_m^{GIS}$ even though it is a gauge-invariant quantity.

Contrary to $\delta_m^{GIS}$, the other gauge invariant perturbation $\delta_m^{GIN}$ has some important attractive features with respect to observability. These are summarized as follows: \begin{itemize} \item It reduces to the Newtonian gauge perturbation $\delta_m^{NG}$ {\it ie} it corresponds to a frame which respects the isotropic expansion of the universe and is therefore more appropriate for description of large scale perturbations. This reduction also simplifies the calculation of this perturbation. \item It drives a scale dependent modification of the Poisson equation for the gauge invariant potential $\phi$. Indeed, the time-time part of the
linearized Einstein equation gives \cite{mukhanov,mabertschinger}
\begin{equation}\label{zero}
\bigtriangledown^2\phi-3\frac{\dot{a}}{a}(\frac{\dot{a}}{a}\psi+\dot{\phi})
=-k^2\phi-3\frac{\dot{a}}{a}(\frac{\dot{a}}{a}\psi+\dot{\phi})
=4\pi G\rho a^2\delta_m^{GIN}.
\end{equation}
Thus, the anticipated scale dependence on scales comparable to the Hubble scale is picked up by the perturbation $\delta_m^{GIN}$. \item It is gauge invariant as anticipated for any observable quantity. \end{itemize}

These features make the gauge invariant $\delta_m^{GIN}$ and the Newtonian gauge $\delta_m^{NG}$ to  which it reduces, an attractive choice for making theoretical calculations to obtain the gravitational potential and the matter density perturbation that can be directly compared with observations on large scales. However, these theoretically obtained quantities need to also be corrected for bias, redshift distortions (due to gravitational potential and peculiar velocities), lensing magnification and volume distortion\cite{Yoo:2009au}.

\section{Conclusion}
I have reviewed the consistency of recent geometric cosmological data with the \lcdm cosmological model.
The standard candle SnIa datasets considered in this study are consistent with \lcdm and with standard rulers at a
level of $95.4\%$ ($2\sigma$) or less for certain prior values of
the matter density $\omm$ in the range $\omm \in [0.25,0.35]$. It is interesting that the ranking sequence based on consistency
with $\Lambda$CDM is practically identical with the corresponding ranking
based on consistency with standard rulers even though the two criteria are completely independent.
Thus, despite the improvement of standard ruler
and standard candle data quality during the last decade the
consistency of $\Lambda$CDM with data has not decreased despite
the fact that $\Lambda$CDM is a simple, specific and well defined
model which appears as a measure-zero point in all generalized
models. On the contrary its consistency seems to be improving with
time as new and more accurate data appear. For example, the
Constitution SnIa dataset which is a very recent compilation with a
drastic improvement on the crucial nearby SnIa sample, is also one
of the most consistent datasets with both $\Lambda$CDM and
standard rulers.

Despite of its excellent consistency with both SnIa standard candles
and CMB-BAO standard rulers, $\Lambda$CDM has to face potential
challenges from other cosmological data
\cite{Perivolaropoulos:2008ud} (e.g. large scale velocity flows,
galaxy and cluster halo profiles, peculiar features of CMB maps
etc.) which may lead the quest for the properties of dark energy
to interesting surprises in the near future. Such surprises may
also come from future standard candle observations or standard
ruler CMB experiments (e.g. Planck \cite{planck}) which are
expected to significantly improve the accuracy of the constraints
discussed in the present study.

I have also discussed dynamical probes of the cosmological constant and reviewed a scale dependent parametrization of the growth rate $f=\frac{d\ln \delta_m}{d \ln a}$ that is free from the sub-Hubble approximation of the standard parametrization (\ref{f0def}). This parametrization described by equation (\ref{newpar}) approximates very well the full linear general relativistic solution up to horizon scales and differs significantly from the standard parametrization on scales larger than about $100 h^{-1}Mpc$. This parametrization describes well the growth of matter density perturbations on large scales in the Newtonian gauge.

Even though the matter density perturbation is a gauge dependent quantity, the Newtonian gauge has certain advantages that make it particularly useful when making a direct comparison of theoretical predictions with observations. These advantages are summarized as follows:
\begin{itemize} \item The evolution of perturbations in the Newtonian gauge picks up the anticipated scale dependence when the Hubble scale is approached. \item The time slicing of the Newtonian gauge respects the isotropic expansion of the background and therefore it is more appropriate to describe large scale fluctuations. On the other hand, the synchronous gauge corresponds to free falling observers at all points and therefore it is physically relevant on small scales where the two gauges give identical evolution of cosmological perturbations. \item Gauge invariant generalizations of the density fluctuation $\delta_m$ and the gravitational potential reduce to the corresponding Newtonian quantities in the Newtonian gauge. This identification with gauge invariant quantities is a prerequisite for the direct observability of the density perturbation and the gravitational potential in the Newtonian gauge \end{itemize}
Even though additional corrections need to be made to the theoretically predicted value of the density perturbation evolution along the lines of Ref. \cite{Yoo:2009au}, it is clear that the Newtonian gauge offers a good starting point for the direct comparison of theoretical predictions of dark energy models with observations.

\end{document}